# Magnetic Domain Wall Motion due to AC Bias-Driven Resonances

Duck-Ho Kim[1†*], Dong-Hyun Kim[2†], Dae-Yun Kim[3,4], Sug-Bong Choe[4], Teruo Ono[1,5], Kyung-Jin Lee[2,6,7], and Se Kwon Kim[8*]

[1]Institute for Chemical Research, Kyoto University, Uji, Kyoto 611-0011, Japan

[2]Department of Semiconductor Systems Engineering, Korea University, Seoul 02841, Republic of Korea

[3]Center for Spintronics, Korea Institute of Science and Technology, Seoul, 02792, Republic of Korea

[4]Department of Physics and Institute of Applied Physics, Seoul National University, Seoul 08826, Republic of Korea

[5]Center for Spintronics Research Network (CSRN), Graduate School of Engineering Science, Osaka University, Osaka 560-8531, Japan

[6]Department of Materials Science & Engineering, Korea University, Seoul 02841, Republic of Korea

[7]KU-KIST Graduate School of Converging Science and Technology, Korea University, Seoul 02841, Republic of Korea

[8]Department of Physics and Astronomy, University of Missouri, Columbia, Missouri 65211, USA

[†]These authors contributed equally to this work.
[*]e-mail: uzes.physics@gmail.com (D.-H. Kim), kimsek@missouri.edu (S. K. Kim)

Most of the existing researches on the dynamics of a domain wall (DW) have focused on the effect of DC biases, where the induced velocity is determined by the bias strength. Here we show that AC biases such as a field or a current are also able to move a DW via synchronization between the DW angle and the phase of the AC bias. The resulting DW velocity is proportional to the driving frequency of the AC bias, but independent of the bias strength, offering potentially low-power operations of DW devices. The AC-bias-driven DW motion is shown to exhibit a phase locking-unlocking transition, a critical phenomenon akin to the Walker breakdown of a DC-bias-driven DW motion. Our work shows that a DW can be driven resonantly by synchronizing its angle to AC biases, shedding a light on hitherto overlooked utility of internal degree of freedom for driving magnetic textures.

The dynamics of topological solitons in magnets has been a topic of long-standing interest because of their intriguing fundamental physics as well as technological applications[1-3]. A prototypical example is a domain wall (DW) in an easy-axis magnet, an interface between two different uniform ground states. In 1974, Schryer and Walker studied the dynamics of a DW induced by an external magnetic field and discovered a novel nonlinear phenomenon, so-called Walker breakdown which refers to the sudden drop of the DW velocity due to the onset of its precessional motion[4]. More recently, current-induced motion of a DW has been studied intensively due to its potential utility as a topologically-protected information carrier in spintronic devices as demonstrated in magnetic DW racetrack memory[3]. Most of the existing researches on the dynamics of a DW have focused on the effects of DC field or current, where the induced velocity is determined by the strength of the external bias, whereas the dynamics of a DW subjected to AC biases has remained as a largely open area except for a few studies on antiferromagnetic DW motion driven by rotating fields[5,6]. The focus of this work is on the AC-bias-induced dynamics of a ferromagnetic DW, which has not been explored yet.

The low-energy dynamics of a DW in a thin ferromagnet strip with perpendicular magnetic anisotropy is known to be well described by the two collective coordinates, its time dependent position $X(t)$ and in-plane angle of the DW's magnetization $\Phi(t)$[4,7,8]. In the previous studies on the DW motion, the primary focus has been on the dynamics of the position variable $X$, whereas the dynamics of the angle variable $\Phi$ has been considered secondary and sometimes undesirable. The origin of this perception can be found in a DW motion driven by a DC field[4]. When the driving field is sufficiently small, the DW moves to decrease the Zeeman energy, but the DW angle $\Phi$ is maintained constant since its dynamics is suppressed by the shape anisotropy. In this regime, there is only one channel for energy dissipation, the dynamics of $X$, and thus all the energy is spent only on moving the DW. When the external field becomes strong enough to overcome the shape anisotropy, the aforementioned Walker breakdown occurs due to the gyrotropic coupling between the position and the angle and the DW precesses and moves simultaneously. The Walker breakdown engenders a new dissipation channel through the angle dynamics $\dot{\Phi}$ compared to the small-field regime, giving rise to the substantial slowdown of the DW. For this reason, finding an efficient way to avoid the Walker breakdown by suppressing the dynamics of the angle $\Phi$ has been a topic of significant interest[9-12].

In this Article, we aim to challenge the conventional subsidiary, often antagonistic, view on the dynamics of the DW angle $\Phi$ by investigating the DW behaviour when its angle is resonantly driven directly by AC biases as a primary control variable. To this end, we study the dynamics of a DW in the simultaneous presence of DC and AC fields. Specifically, we consider the DC magnetic field that is sufficiently strong to put the system above the Walker breakdown so that the DW angle keeps changing at a certain precession frequency, and investigated the effect of the AC magnetic field on the DW velocity. We find that the DW velocity changes linearly as a function of the frequency of the AC field, not the strength of it, when the driving frequency is close to the precessional frequency of the DW. The frequency-dependent DW velocity is explained by the phase locking between the AC field and the DW precession, which is analogous to the phase locking between interacting oscillators[13]. We also discover a critical phenomenon in the AC-field-driven DW motion, which is identified as a phase locking-to-unlocking transition.

In addition, we find the same phenomenon in the current-induced DW motion in the presence of DC spin-transfer torque (STT) and AC spin-orbit torque (SOT), which can be realized in heterostructures consisting of a metallic ferromagnet and a heavy metal with insulating barrier, e.g., Pt/NiO/CoTb used in Ref. 14. The result AC-current-induced DW motion is independent of the DW type differing from the field-driven case, and thus can be used for realizing DW racetrack memory[3]. All the theoretical results are supported by numerical simulations. We envision that hitherto overlooked internal degrees of freedom of topological magnetic solitons can serve as alternative handles for the dynamics of the solitons, wherein AC biases can be used as useful and versatile tools.

Our model system is a thin ferromagnetic film with perpendicular magnetic anisotropy, which can be described by the following Hamiltonian: $U = \int dx [(A(\partial_x \boldsymbol{m})^2 - K m_z^2 + K_y m_y^2)/2 - M_S \boldsymbol{H} \cdot \boldsymbol{m}]$, where $\boldsymbol{m} = (m_x, m_y, m_z)$ is the unit vector in the magnetization direction, $A > 0$ is the exchange coefficient, $K > 0$ is the easy-axis anisotropy perpendicular to the film which includes the shape anisotropy for a thin film, $K_y > 0$ is the hard-axis anisotropy, $M_S$ is the saturation magnetization, and $\boldsymbol{H}$ is an external field. The solution for a DW connecting the two ground states, $\boldsymbol{m}(x \to \pm\infty) = \pm q\hat{\boldsymbol{z}}$, is given by $\boldsymbol{m} = (\text{sech}[(x-X)/\lambda]\cos\Phi, \text{sech}[(x-X)/\lambda]\sin\Phi, q\tanh[(x-X)/\lambda])$ [4], where $\lambda = \sqrt{A/K}$ parametrizes the DW width and $q = \pm 1$ represents the type of the DW. Here, $X$ is the position of the DW, which represents a zero-energy mode associated with the spontaneous breaking of the translational symmetry by the DW; $\Phi$ is the in-plane angle of the DW's magnetization, which represents an internal degree of freedom. The low-energy dynamics of the DW can be described with these two collective coordinates, $X(t)$ and $\Phi(t)$.

The dynamics of a ferromagnet can be described by the Landau-Lifshitz-Gilbert (LLG) equation[15,16]: $\dot{\boldsymbol{m}} - \alpha \boldsymbol{m} \times \dot{\boldsymbol{m}} = -\gamma \boldsymbol{m} \times \boldsymbol{H}_{\text{eff}}$, where $\gamma > 0$ is the gyromagnetic ratio, $\boldsymbol{H}_{\text{eff}} = -\delta U/(M_S \delta \boldsymbol{m})$ is the effective magnetic field, and $\alpha > 0$ is the Gilbert damping constant parametrizing the energy-dissipation rate through the magnetic dynamics. For the external field, we consider both a DC field along the z axis and an AC field along the x axis: $\boldsymbol{H} = H_z \hat{\boldsymbol{z}} + H_x \cos(\omega t) \hat{\boldsymbol{x}}$. See Fig. 1a for the schematic illustration of a DW and external fields. Without loss of generality, we can consider the cases with $\omega > 0$, $H_x > 0$, and $H_z > 0$. The equations of motion for $X$ and $\Phi$ can be obtained from the LLG equations within the collective coordinate approach[7,8]. One for the angle is given by

$$\left(\alpha + \frac{1}{\alpha}\right)\dot{\Phi} = \frac{\gamma H_z}{\alpha} + \frac{\gamma H_K \sin(2\Phi)}{4} + \frac{\pi \gamma H_x \cos(\omega t) \sin \Phi}{2}, \qquad (1)$$

where $H_K \equiv 2K_y/M_S$ is the hard-axis anisotropy field. The equation of motion for the position is given by

$$\dot{X} = q\lambda/\alpha \left(\dot{\Phi} - \gamma H_z\right). \qquad (2)$$

Note that the DW velocity $\dot{X}$ depends linearly on the DW angle precession $\dot{\Phi}$ in Eq. (2), which is the main property of the equation that is utilized in this work. The first line on right-hand side of Eq. (1) shows that the DC field $H_z$ tends to rotate the DW, and that can be compensated by the hard-axis anisotropy field $H_K$. Here, we consider the cases where the DC field is sufficiently strong so that it dominates the hard-axis anisotropy and thereby induces the DW precession: $H_z > \alpha H_K/4$, which is above the so-called Walker breakdown field[4]. In this regime, for the long-term dynamics, the time-averaged precession is finite, $\langle \dot{\Phi} \rangle > 0$, and thus the hard-axis anisotropy term $\propto H_x$ in Eq. (1) that is sinusoidal in the angle can be neglected over the other terms[7,8]. By taking this approximation, Eq. (1) can be recast into

$$\dot{\Phi} = \omega_0 + \omega_1 \sin(\Phi - \omega t) + \omega_1 \sin(\Phi + \omega t), \qquad (3)$$

where $\omega_0 \equiv \gamma H_z/(1 + \alpha^2)$ is the DW rotation frequency in the absence of the AC field and $\omega_1 \equiv \pi \alpha \gamma H_x/(4(1 + \alpha^2))$ represents the magnitude of the AC field. The obtained equation without the last term is called the Adler equation which is known to have a steady-state phase-locked solution $\dot{\Phi} = \omega$ in the studies about the injection locking of a nonlinear oscillator coupled to a resistance-inductor-capacitor circuit[13]. The Adler equation has been invoked in spintronics to explain the phase locking of the spin-torque oscillator to an AC current[17-19].

When the driving frequency $f = \omega/2\pi$ is sufficiently close to the precession frequency $f_M = \omega_0/2\pi$ of the magnetization of the DW $M_{DW}$ (see Fig. 1**b**), the criteria for which will be determined below, the DW angle is synchronized with the AC field (i.e. the resonance between the DW angle dynamics and an AC field). Between the two possible simplest choices for synchronization, $\dot{\Phi} \approx \omega$ and $\dot{\Phi} \approx -\omega$, the first term $\omega_0 > 0$ in Eq. (3) drives the system into the former synchronization. In this phase-locked regime, the DW angle satisfies $\Phi(t) = \omega t + \delta\Phi(t)$ with $\langle \delta\dot{\Phi} \rangle = 0$, where $\langle \cdots \rangle$ denotes the averaging over long time (compared to the frequency $\omega$). From Eq. (2), the resultant time-averaged DW velocity $V = \langle \dot{X} \rangle$ is given by

$$V = \frac{q\lambda}{\alpha}\omega - \frac{q\lambda}{\alpha}\gamma H_z. \tag{4}$$

This is our first main result: the DW velocity in the simultaneous presence of DC and AC fields in the phase locked regime. The first term $\propto \omega$ is the DW velocity induced by the AC field by the synchronization of the DW angle and the AC field. Note that it is a linear function of the frequency of the AC field and it is independent of the magnitude of the AC field. The second term $\propto H_z$ is the well-known DC-field-induced DW velocity below the Walker breakdown[4].

The necessary criteria to be in the phase-locked regime can be found as follows. In terms of $\delta\Phi$, Eq. (3) reads $(\omega - \omega_0) + \delta\dot{\Phi} = \omega_1 \sin(\delta\Phi) + \omega_1 \sin(\delta\Phi + 2\omega t)$. If there were no second term, then $\delta\Phi = \arcsin\left[\frac{(\omega-\omega_0)}{\omega}\right]$ solves the equation, which yields the phase-locked condition $|\omega - \omega_0| < \omega_1$. However, the last term necessitates the inclusion of the sinusoidal contributions, changing the obtained condition. To obtain the self-consistency condition for the phase-locked regime, we use an ansatz $\delta\Phi(t) = \delta\Phi_0 + A\cos(2\omega t) +$ $B\sin(2\omega t)$ with the time-independent $\delta\Phi_0$ and solve the equation for $\delta\Phi_0$, $A$, and $B$ to linear order in $A$ and $B$ by assuming $|\omega_1/\omega_0| \ll 1$. The resultant critical frequencies, whose derivation is in Supplementary Information, are given by

$$\omega_c^{u,l} = \frac{1}{2}\left(\omega_0 \pm \omega_1 + \sqrt{\omega_0^2 \pm 2\omega_0\omega_1}\right), \tag{5}$$

where $+$ and $-$ are for the upper $\omega_c^u$ and the lower $\omega_c^l$ critical frequencies, respectively. This is our second main result. Note that the strength $H_x$ of the AC field determines the frequency range where the obtained phase-locking DW velocity $V$ [Eq. (4)] can be realized.

Our analytical results are compared with micromagnetic simulations. See Methods for the details of the simulation configurations. We present the results for the DW with $q = 1$ below; The simulation results with $q = -1$ show the analogous agreement with the theory (not shown). Figure 2**a** shows the results for the DW velocity $V$ as a function of the AC-field frequency $f = \omega/2\pi$ for the DC field $H_z = 5$ mT and the AC field $H_x = 40$ mT. The plot shows that DW velocity changes linearly as a function of the frequency of the AC field when the driving frequency is close to the precessional frequency of the DW: the frequency of the DW $f_M$ is same as $f$ (see Fig. 2**c**). The frequency-dependent DW velocity is explained by the phase locking between the AC field and the DW precession, which is analogous to the phase locking between interacting oscillators[13]. The purple solid line shows the analytical result for $V$ [Eq. (4)] in the phase-locked regime. Figure 2**b** shows the simulation results in the wider range of $f$. The plot clearly shows the phase locking-to-unlocking transition. The red (blue) box represents the phase-locked (phase-unlocked) regime. When the driving frequency is far away from the precessional frequency of the DW, the velocity is no longer controlled by the AC-field frequency: $f_M$ does not follow $f$ (see Fig. 2**d**).

Figure 3**a** and 3**b** show the simulation results (symbols) for $H_z =$ 5 and 20 mT, respectively, with various AC fields $H_x =$ 5, 10, 20, and 40 mT. The analytical results shown as the purple solid lines and the numerical results agree well. Note that the DW velocity is independent of the magnitude of the AC field $H_x$ within the phase-locked regime. Figure 3**c** shows the numerical results (symbols) and the analytical result (solid lines) [Eq. (5)] for the upper and the lower critical frequencies $f_c^{u,l}$ $(\equiv \omega_c^{u,l}/2\pi)$ for the phase locking-unlocking transition with various $H_z$ and $H_x$, showing good agreement as well. The extra set of simulation results, which are obtained for different material parameters, and their agreement with the analytical results are discussed in Supplementary Information.

For practical applications of a DW as a memory unit in spintronics, it is important to achieve the current-induced unidirectional DW motion without relying on an external field[3]. For this reason, we investigate the current-induced DW motion in the simultaneous presence of a DC STT and an AC SOT. The possible experimental setup is shown in Fig. 4**a**, consisting of a metallic magnet and a heavy metal with an insulating barrier. One readily available material platform is Pt/NiO/CoTb heterostructure, where Pt has been shown to be able to exert SOT on CoTb through an antiferromagnetic insulator NiO[14].

The dynamics of a ferromagnet in the presence of STT[20,21] and SOT[22-26] can be described by the augmented LLG equation: $\dot{\boldsymbol{m}} - \alpha \boldsymbol{m} \times \dot{\boldsymbol{m}} = -\gamma \boldsymbol{m} \times \boldsymbol{H}_{\text{eff}} + \boldsymbol{\tau}_{\text{STT}} + \boldsymbol{\tau}_{\text{SOT}}$, where DC STT is given by $\boldsymbol{\tau}_{\text{STT}} = P(\boldsymbol{J}_{\text{DC}} \cdot \nabla)\boldsymbol{m} - \beta P \boldsymbol{n} \times (\boldsymbol{J}_{\text{DC}} \cdot \nabla)\boldsymbol{n}$ and AC SOT is given by $\boldsymbol{\tau}_{\text{SOT}} = \gamma B_{\text{F}} \cos(\omega t)\, \boldsymbol{m} \times \hat{\boldsymbol{y}} + \gamma B_{\text{D}} \cos(\omega t)\, \boldsymbol{m} \times (\boldsymbol{m} \times \hat{\boldsymbol{y}})$. Here, $\boldsymbol{J}_{\text{DC}} = J_{\text{DC}}\hat{\boldsymbol{x}}$ is the DC charge current density, $\beta$ is the dimensionless number parametrizing the dissipative component of STT, $P = (\hbar\gamma/2eM_{\text{S}})(\sigma_\uparrow - \sigma_\downarrow)/(\sigma_\uparrow + \sigma_\downarrow)$ (with $e > 0$ the electron charge) represents the polarization of the spin-dependent conductivity $\sigma_s$ (s =↑ chosen along $-\boldsymbol{m}$), $B_{\text{F}} = (\hbar/2e)(\theta_{\text{F}}J_{\text{AC}}/M_{\text{S}}t_z)$ and $B_{\text{D}} = (\hbar/2e)(\theta_{\text{D}}J_{\text{AC}}/M_{\text{S}}t_z)$ are, respectively, the field-like and the damping-like SOT magnitudes induced by the AC current density $J_{AC} = J_{AC}\cos(\omega t)\hat{x}$, $t_z$ is the thickness of the ferromagnet, and $\theta_F$ and $\theta_D$ are the spin Hall angles for the corresponding SOT components. The equations of motion for $X(t)$ and $\Phi(t)$ can be obtained within the collective coordinate approach[7,8]. The equation for the angle is given by

$$\left(\alpha + \frac{1}{\alpha}\right)\dot{\Phi} = \frac{q(\beta-\alpha)PJ_{DC}}{\alpha\lambda} + \frac{\gamma H_K \sin(2\Phi)}{4} - \frac{\pi\gamma}{2}\left(\frac{B_D}{\alpha} + B_F\right)\cos(\omega t)\cos\Phi, \qquad (6)$$

The equation of motion for $X$ is given by

$$\dot{X} = \frac{q\lambda}{\alpha}\left(\dot{\Phi} - \frac{q\beta PJ_{DC}}{\lambda} + \frac{\pi}{2}\gamma B_D \cos(\omega t)\cos\Phi\right), \qquad (7)$$

We consider the cases where the DC STT is sufficiently strong to put the DW dynamics above the Walker breakdown, $(\beta - \alpha)PJ_{DC}/(\alpha\lambda) > \gamma H_K/4$, so that the DW precesses with the finite average angular velocity, $\langle\dot{\Phi}\rangle > 0$. Then, the phase-locked solutions to the above two equations for the current-induced DW motion can be obtained in an analogous way for the solutions to Eqs. (1) and (2). In the phase-locked regime, where $\langle\dot{\Phi}\rangle = \omega$, the DW velocity is given by

$$V = \frac{\lambda(B_F - \alpha B_D)}{B_D + \alpha B_F}\omega - \frac{B_D + \beta B_F}{B_D + \alpha B_F}PJ_{DC}. \qquad (8)$$

This is our third main result: the DW velocity in the simultaneous presence of a DC STT and an AC SOT in the phase-locked regime. The first term $\propto \omega$ is the DW velocity induced by the synchronization of the DW angle dynamics and the AC SOT, which is linear in the frequency and independent of the magnitude. Note that the DW velocity is independent of the DW type $q = \pm 1$, which is in contrast to the field-induced DW velocity [Eq. (4)]. This has an important consequence for practical applications: A train of DWs of different types of $q = \pm 1$ can be driven all together in the same direction, which is crucial to realize the magnetic DW racetrack memory[3]. The necessary condition for the DW motion to be in the phase-locked regime can be obtained by Eq. (5) analogously to the field-driven case.

The analytical results are checked with numerical simulations. The results are shown in Fig. 4**b**, where the DC current density $J_{DC} = 2\times10^{12}$ A/m$^2$ and the AC current densities $J_{AC} = 2, 4, 6, 8\times10^{10}$ A/m$^2$ are used. The symbols and the lines show the numerical and the analytical results [Eq. (8)], showing good agreement. Fig. 4**c** shows the numerical and the analytical result [Eq. (5)] for the critical frequencies $f_c^{u,l}$ for the phase locking-unlocking transition, which also exhibits good agreement except for the frequency close to 0 where the high-frequency approximation for Eq. (5) becomes invalid. We would like to mention that, in the simulations, the DC current of the large magnitude is applied for illustrative purpose so that the DW prcession frequency $f_M$ is high enough to exhibit both the upper and the lower critical breakdowns. However, the AC-biased DW motion are expected to work as long as the DW is in the precessing regime which can be achieved by the DC current of much smaller magnitudes.

We have shown analytically and numerically that, in the simultaneous presence of a DC and an AC bias, where biases can be either external fields or currents, a ferromagnetic DW can synchronize its angle with the AC bias, which in turn gives rise to its motion via the gyrotropic coupling between the angle and the position. One notable feature of the resultant velocity is that it is linearly proportional to the frequency of the AC bias and independent of its magnitude. We have also discussed the transition from phase-locking to phase-unlocking, and obtained the necessary criteria for the phase-locking behaviour. In this work, we have focused on the cases where the DC bias is strong enough to induce the precessional dynamics of the DW (i.e., above the Walker breakdown). Further investigations are needed to understand the DW dynamics in the presence of general DC and AC biases and thereby obtain the full non-equilibrium phase diagram of the DW motion. In addition, we expect that AC-bias-driven DW motion exhibits fractional synchronization phenomena where the angle precession frequency is locked to a non-integral rational multiple of the driving frequency, which we leave as a future research topic. Lastly, the identified mechanism for the AC-bias-driven DW motion utilizes the gyrotropic coupling between the DW angle and the position of ferromagnetic DWs, and thus it would not be operative for antiferromagnetic DWs, which lack in the gyrotropic coupling as manifested by the absence of the Walker breakdown[27,28]. We hope that our work will trigger new theoretical and experimental researches in the future, where the DW angle is viewed as an active degree of freedom for controlling the DW.

**Methods**

**Micromagnetic simulations.** For the micromagnetic simulations, the following material parameters are used: $M_S = 10^6$ A/m, $A = 10^{-11}$ J/m, $K = 9 \times 10^5$ J/m³, $K_y = 500$ J/m³, $\alpha = 0.02$, and $\gamma = 1.76 \times 10^{11}$ rad/s·T. The considered sample geometry (length × width × thickness) is 3200 nm × 50 nm × 1 nm with the cell size of 0.4 nm × 50 nm × 1 nm. The dipolar interactions are accounted for by including the local demagnetizing field. The parameters for STT and SOT are $P = 0.5(\hbar\gamma/2eM_S)$, $\beta = 0.06$, $\theta_F = 0.28$, and $\theta_D = -0.07$.

**Figure Captions**

**Figure 1. Domain wall (DW) subjected to a DC and an AC fields. a**, Schematic for a DW motion in a ferromagnet with perpendicular magnetic anisotropy subjected to a DC field $H_z$ (along the z axis) and an AC field $H_x \cos(\omega t)$ (along the x axis). Blue area, magnetization-up state (circled dot); red area, magnetization-down state (crossed circle). **b**, Schematic for an angular motion of the magnetization of the DW and an AC field rotating at the frequencies $f_M$ and $f$, respectively.

**Figure 2. DW angular and linear motion in the phase-locked and the phase-unlocked regimes. a**, and **b**, show the DW velocity $V$ as a function of the frequency $f$ of the AC field in the presence of the DC field, $H_z$ =5 mT and $H_x$ =40 mT. The phase-locking regime (red area) and phase-unlocking regime (blue area) were shown in Fig. 2**b**. **c**, and **d**, illustrate the time dependence of a DW in-plane magnetization $M_{DW}$ and an AC field $H_x$ in the phase-locked regime and unlocked regime, respectively.

**Figure 3. DW velocity induced by the resonance between the DW angle and the AC field.** **a**, and **b**, show the DW velocity $V$ as a function of the frequency $f$ of the AC field in the presence of the DC field, $H_z$ =5 and 20 mT, respectively. The symbols and the lines are from numerical simulations and analytical solutions [Eq. (6)], respectively. Insets of Fig. 3**a** and **b** are the simulation results in the wider range of $f$. **c**, The upper and the lower critical frequencies $f_c^{u,l} \left( \equiv \omega_c^{u,l}/2\pi \right)$ for the phase locking-to-unlocking transition as a function of the AC-field magnitude. The symbols and the lines are from numerical simulations and analytical solutions [Eq. (8)], respectively.

**Figure 4. DW velocity induced by the resonance between the DW angle and the AC current. a**, Schematics for the current-induced DW motion, where a ferromagnet is subjected to STT by a DC current density $J_{DC}$ through the magnetic layer and SOT by an AC current density $J_{AC}\cos(\omega t)$ through a heavy-metal layer. **b**, The DW velocity $V$ as a function of the frequency $f$ of the AC current. The symbols and the lines are from numerical simulations and analytical solutions [Eq. (12)], respectively. **c**, The upper and the lower critical frequencies $f_c^{u,l}$ for the phase locking-unlocking transition. The symbols and the lines are from numerical simulations and analytical solutions [Eq. (8)], respectively.

**Acknowledgements**

S.K.K. was supported by the startup fund at the University of Missouri. This work was supported by the JSPS KAKENHI (Grant Numbers 15H05702, 26870300, 26870304, 26103002, 26103004, 25220604, and 2604316), Collaborative Research Program of the Institute for Chemical Research, Kyoto University, and R & D project for ICT Key Technology of MEXT from the Japan Society for the Promotion of Science (JSPS). This work was partly supported by The Cooperative Research Project Program of the Research Institute of Electrical Communication, Tohoku University. D.H.K. was supported as an Overseas Researcher under the Postdoctoral Fellowship of JSPS (Grant Number P16314). K.J.L. was supported by the National Research Foundation of Korea (NRF-2017R1A2B2006119) and the KIST Institutional Program (Project No. 2V05750). D.Y.K. and S.B.C. were supported by the Samsung Science & Technology Foundation (SSTF-BA1802-07) and the National Research Foundations of Korea (NRF) funded by the Ministry of Science, ICT (MSIT) (2015M3D1A1070465). D.Y.K. was supported by the Korea Institute of Science and Technology (KIST) institutional program (No. 2E29410) and the National Research Council of Science & Technology (NST) grant (No. CAP-16-01-KIST) funded by the Korea government (Ministry of Science and ICT).

**Author contributions**

D.-H.K. and S.K.K. conceived the idea, D.-H.K., K.-J.L., and S.K.K. planned and designed the study. D.-H.K., Dong-Hyun. K., D.-Y.K., and K.-J.L. performed the micromagnetic simulation and the analysis. S.K.K. supported the theory. D.-H.K., K.-J.L., and S.K.K. wrote the manuscript. All authors discussed the results and commented on the manuscript.

**Additional information**

Correspondence and request for materials should be addressed to D.H.K. and S.K.K.

**Competing financial interests**

The authors declare no competing financial interests.

**Data and materials availability**

All data are available in the main text or the supplementary information.

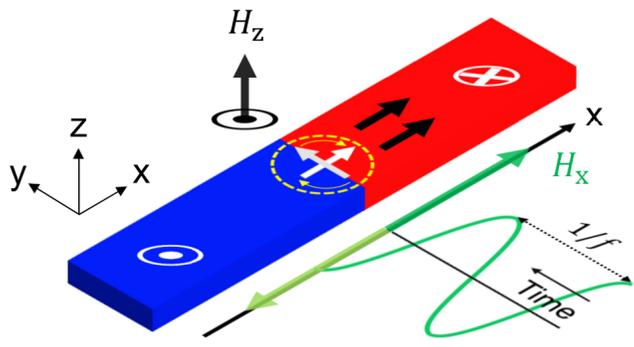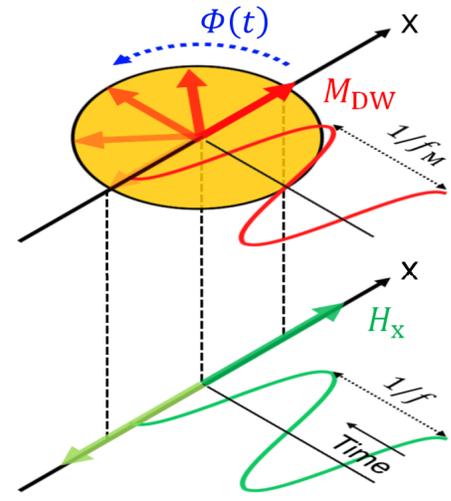

Figure 1

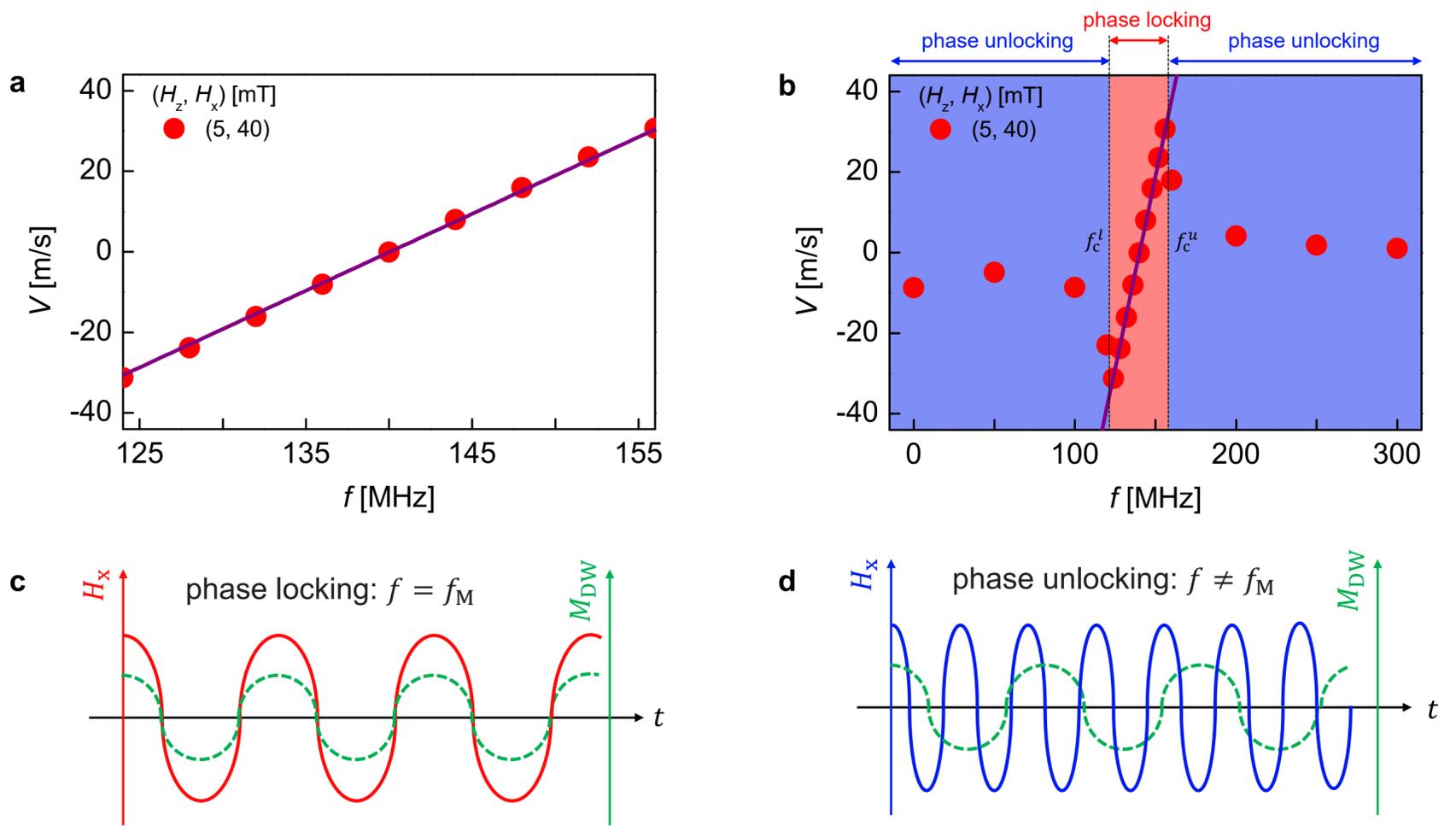

Figure 2

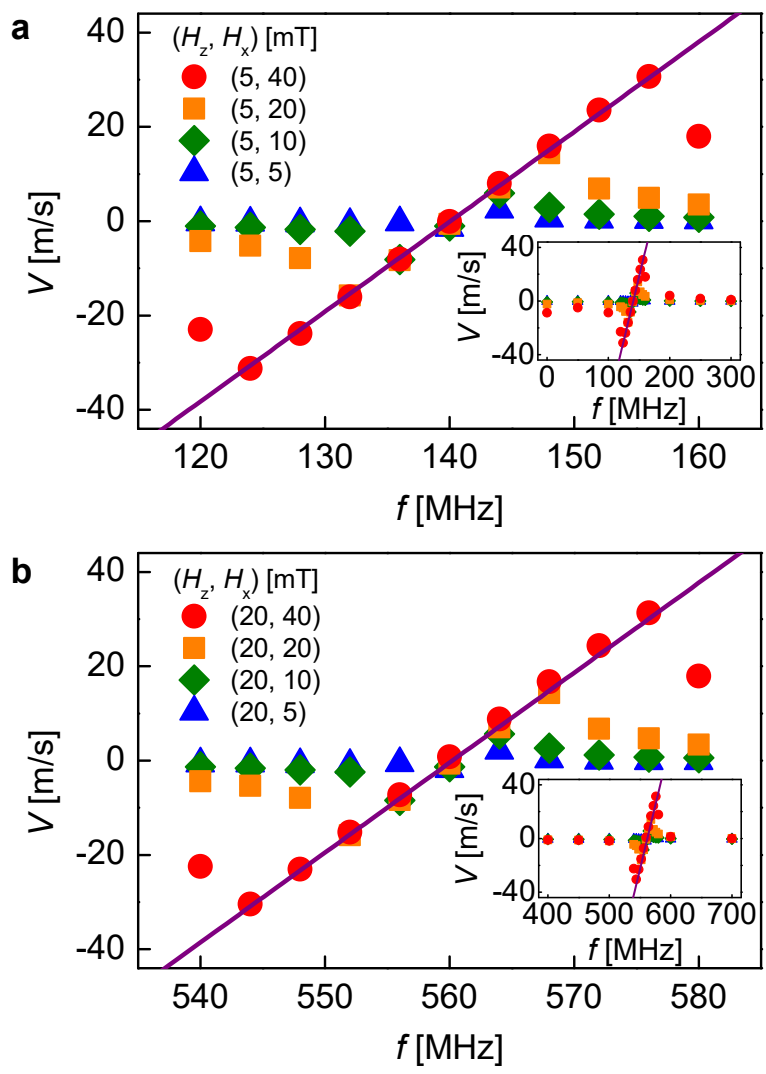
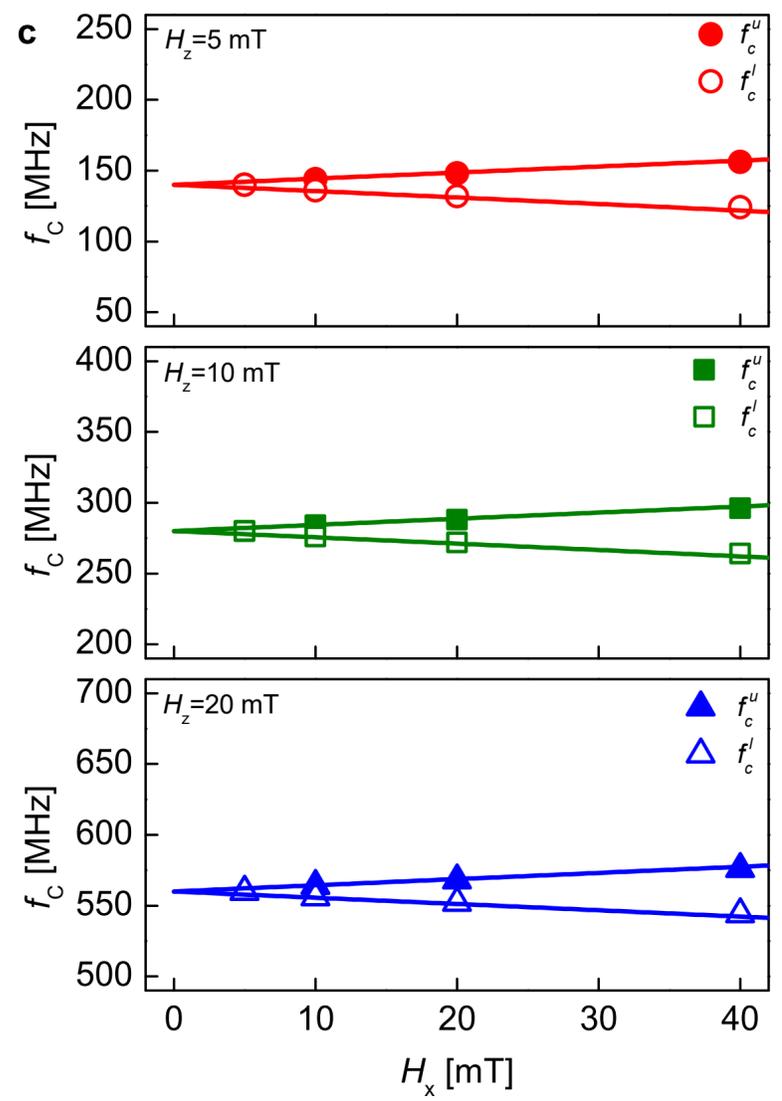

Figure 3

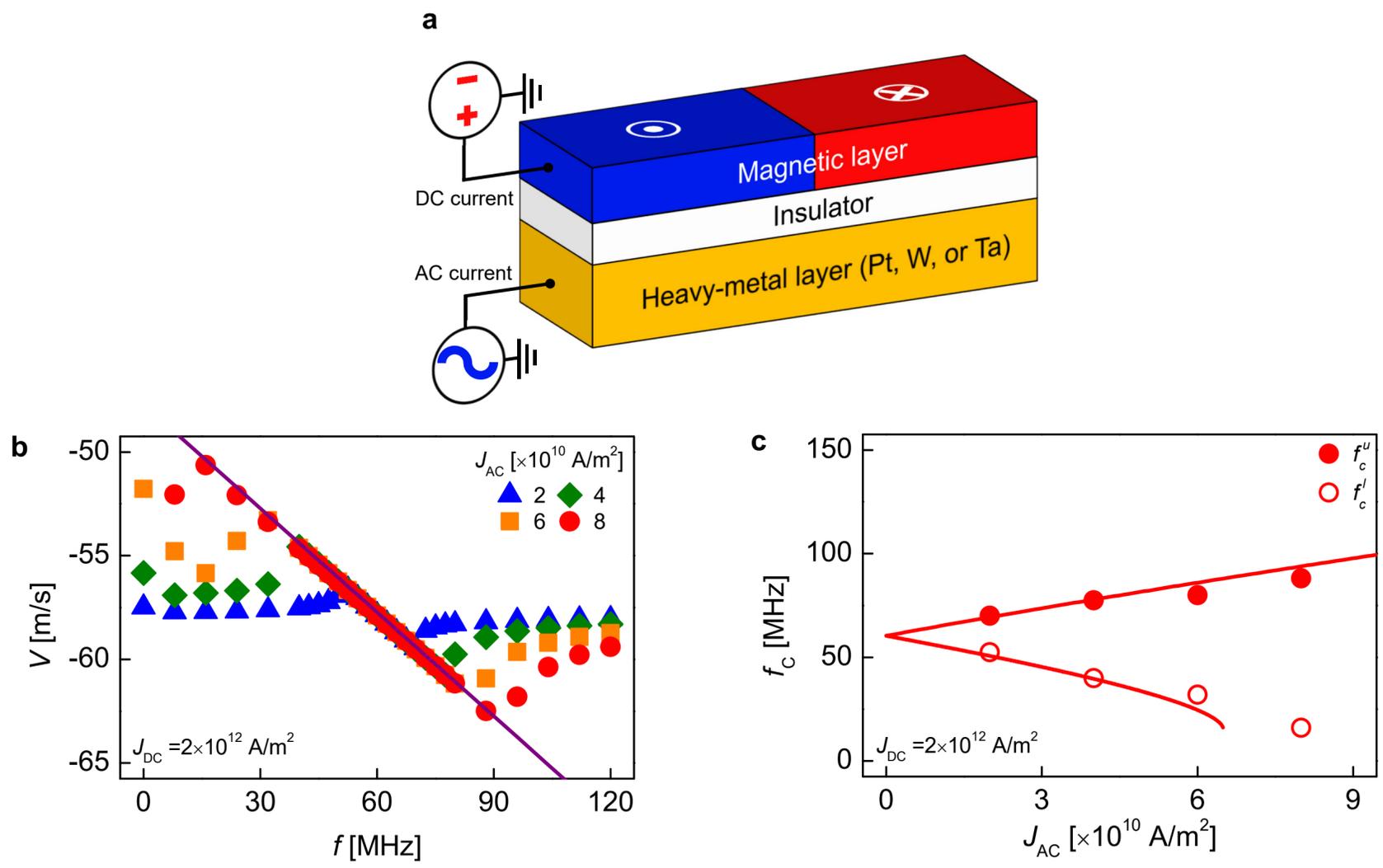

**Figure 4**